\documentclass[aps,prc]{revtex4}        
\usepackage{graphicx}
\begin{document}
\draft
\title{Potential renormalization by tunneling induced intrinsic transition}
\author{Noboru Takigawa$^{a,c}$,
\thanks{E-mail address: takigawa@nucl.phys.tohoku.ac.jp}
Sachie Kimura$^{a,c}$
\thanks{E-mail address: sachie@nucl.phys.tohoku.ac.jp}
and David M. Brink$^{c}$
\thanks{E-mail address: brink@thphys.ox.ac.uk}
}
\address{$^{a}$Department of Physics, Tohoku University, 
Sendai 980-8578, Japan\\ 
$^{c}$Laboratorio Nazionale del Sud, Istituto Nazionale di Fisica Nucleare,
via S. Sofia 44, I-95123 Catania, Italy
}
\date{\today}


\begin{abstract}
The screening effects by bound target electrons in low energy 
nuclear reactions in laboratory experiments can be well represented 
in terms of a constant shift of the tunneling potential barrier 
as postulated in all previous analyses if the electronic 
cloud were static or structureless and lied outside the tunneling region.
In this paper, using a semiclassical mean field analysis of a simple 
two-level coupled-channels problem, we suggest that 
there could occur a strong quantum transition of electrons from excited 
to the ground states in the tunneling region, leading to a 
spatially dependent screening potential. 
We thus show the necessity of the dynamical treatment of the tunneling 
region in order to properly assess the screening effects by 
bound target electrons in low energy nuclear reactions in laboratories.
\end{abstract}

\maketitle

\section{introduction}
The rate of nuclear reactions at extremely low energies 
in laboratory experiments shows a marked enhancement 
over the extrapolation from high energy data\cite{SUMRolfsSomorjai}.
This phenomenon has been attributed to the screening effect 
by the bound electrons in the target nucleus, 
and has been discussed in many papers for almost a decade.  
Yet, a clear understanding is missing. 
A standard approach is to analyze the data by assuming that the 
electrons lead to a constant potential shift in the tunneling 
region, which is called the screening energy. 
A surprise is that the phenomenological 
value of the screening energy thus obtained exceeds the so called 
adiabatic limit, which is conventionally given by the difference between the 
binding energies in the target and the united atoms and 
is thought to give the maximum screening energy, for all systems 
so far studied. One should, however, be aware of the fact that a 
spatially constant potential renormalization can be justified only for a 
static 
electronic cloud, namely, when the electronic state 
makes no transition in the tunneling region. 

The screening problem can be understood as an example of the 
so called macroscopic quantum tunneling, where intrinsic degrees 
of freedom affect the tunneling probability of a macroscopic variable, 
by considering the electrons and the relative motion between the 
projectile and target nuclei as the intrinsic and macroscopic degrees 
of freedom, respectively. 
In this paper, we show that a strong intrinsic transition can happen 
to facilitate the tunneling probability leading to a novel, spatially 
dependent potential renormalization. In order to demonstrate this, 
we use a simple two-level coupled-channels problem and analyze the 
associated tunneling process based on a semi-classical mean field 
theory, which naturally describes the effects of intrinsic degrees 
of freedom on the quantum tunneling of a macroscopic variable in terms 
of a potential renormalization. 

The paper is organized as follows. 
In sect.2, we briefly 
describe the essence of the semi-classical mean field theory for 
quantum tunneling and apply it to a two-level coupled-channels problem. 
In sect.3, we examine the accuracy of the semi-classical mean field theory 
through the comparison of the barrier transmission probability 
given by the semiclassical mean field theory with that obtained by 
a direct numerical solution of the coupled-channels equations. 
In sect.4, we discuss the properties of the transition of the 
intrinsic state and the associated potential renormalization.
We summarize the paper in sect.5.

\section{Semiclassical mean field theory for quantum tunneling}

\subsection{Basic coupled-equations}

Our interest is to study the time evolution of a system, which 
consists of a macroscopic variable $X$ which undergoes a 
quantum tunneling and intrinsic degrees of freedom $\xi$. 
In the screening problem, the former and the latter correspond 
to the coordinate of the relative motion between 
the projectile and target nuclei 
and electronic degrees of freedom, respectively, if one 
ignores the intrinsic structure of nuclei. We assume that 
the time dependent Schr{\"o}dinger equation for the total system 
is given by,
\begin{eqnarray}
i\hbar \frac{\partial \Psi(X,\xi,t)}{\partial t}=
  \left[H_N(X)+H_0(\xi)+V(X,\xi)\right]\Psi(X,\xi,t)
\end{eqnarray}
where we used the notation $H_N(X)$ for 
$-{{\hbar^2}\over{2\mu}}{{\partial^2}\over{\partial X^2}}+U(X)$,
$U(X)$ being the bare interaction between the projectile and target 
nuclei. The $H_0(\xi)$ and $V(X,\xi)$ represent the unperturbed 
Hamiltonian of the electrons and the interaction between the 
electrons and nuclei. 

We now approximate $\Psi(X,\xi,t)$ by a product 
of the wave functions for the macroscopic and intrinsic motions
\begin{eqnarray}
\Psi(X,\xi,t)=\phi(\xi,t)\eta(X,t)
\end{eqnarray}
and determine the wave function in each space 
based on the following variational principle,
\begin{eqnarray}
\frac{\delta {\cal L}}{\delta \phi^*(\xi,t)} &=& 0 \\
\frac{\delta {\cal L}}{\delta \eta^*(X,t)} &=& 0
\end{eqnarray}
where the Lagrangian is defined by
\begin{eqnarray}
{\cal L}=\int_{t_0}^T dt~ 
  \langle \phi \eta \vert (i\frac{d}{dt}-H)\vert \phi \eta \rangle
\end{eqnarray}
We call this approach a mean field theory or a Hartree approximation 
(see \cite{tkbs00,tkb01a} for details).

One can try to numerically solve the resultant time dependent 
coupled-equations for $\phi(\xi,t)$ and $\eta(X,t)$. 
However, as is well known \cite{gri80}, that approach 
is adequate to describe only the classically allowed region. 
It suffers from the so-called spurious correlation effect to handle 
the tunneling region. As a practical method to circumvent this problem, 
we treat the macroscopic motion 
classically and resort to the imaginary time approach 
to describe the tunneling process. Identifying the classical 
variables with the expectation values of the corresponding 
operators at each time, we have the 
following coupled equations for the classically allowed region
\begin{eqnarray}
i\hbar \frac{\partial \phi(\xi,t)}{\partial t}=
\left[ H_0(\xi)+V(X(t),\xi)\right] \phi(\xi,t), 
\label{envc}
\end{eqnarray}
\begin{eqnarray}
\mu\frac{\partial^2}{\partial t^2}X(t)=-\frac{\partial}{\partial X}
\big[U(X)+\frac{<\phi(t)\vert \big[H_0+V(X(t),\xi)\big] \vert\phi(t)>}
{<\phi(t)\vert \phi(t)>}
\big].
\label{relc} 
\end{eqnarray}
They lead to the following energy conservation law
\begin{eqnarray}
\frac{1}{2}\mu(\frac{\partial}{\partial t}X(t))^2
+U(X(t))+\frac{<\phi(t)\vert 
\big[H_0+V(X(t),\xi)\big]\vert\phi(t)>}
{<\phi(t)\vert \phi(t)>}=E.
\label{energycc} 
\end{eqnarray}
where the total energy E is given by the sum 
of the initial kinetic energy $E_K$ and the initial 
intrinsic energy $\epsilon_0$.

As already mentioned, we describe the tunneling process using the imaginary 
time $t=-i\tau$. 
We must, however, find also the proper way to describe the effects of 
intrinsic motion during the tunneling process. For example, it is not 
obvious whether the potential renormalization is given by the 
same expression as that in eq.(\ref{relc}) or one should replace 
$<\phi(t)\vert$ by the adjoint state. 
We discuss in a separate paper \cite{tkb01a} the 
effective potential equivalent to the dynamical norm factor 
which takes non adiabatic effect into account \cite{tak95}
and show that the proper generalization of 
eqs.(\ref{envc}) through (\ref{energycc}) 
to the tunneling region are given by  
\begin{eqnarray}
& & \hbar \frac{\partial \phi(\xi,\tau)}{\partial \tau}=
-\left[ H_0(\xi)+V(X(\tau),\xi)\right] \phi(\xi,\tau). 
\label{envt}\\
& & \mu\frac{\partial^2}{\partial \tau^2}X(\tau)=\frac{\partial}{\partial X}
\big[U(X)+\frac{<\phi(\tau)\vert \left[
H_0(\xi)+V(X,\xi)\right]  \vert\phi(\tau)>}
{<\phi(\tau)\vert \phi(\tau)>}\big],
\label{relt}\\ 
& & -\frac{1}{2}\mu(\frac{\partial}{\partial \tau}X(\tau))^2
+U(X(\tau))+\frac{<\phi(\tau)\vert 
\big[H_0+V(X(\tau),\xi)\big]\vert\phi(\tau)>}
{<\phi(\tau)\vert \phi(\tau)>}=E,
\label{energyct} 
\end{eqnarray}
We see in eqs.(\ref{relc}) and (\ref{relt}) 
the explicit expression of the potential 
renormalization in the classical and tunneling regions, respectively.

\subsection{Application to a two-level coupled-channels problem}

We now apply the semi-classical mean field theory to a 
coupled-channels problem, which has proved to be very powerful to discuss, 
e.g., the effects of nuclear intrinsic degrees of freedom on the 
cross section of heavy-ion fusion reactions at energies below 
the Coulomb barrier \cite{bt98}. Here we consider a simple model, 
where the intrinsic motion has two levels
\begin{equation}
  H_0 |\phi_m \rangle = \epsilon_m |\phi_m \rangle 
  \hskip 2cm (m=1,2), 
\end{equation}
and the coupling Hamiltonian is separable
\begin{equation}
V(X,\xi) = f(X) \cdot {\hat M}(\xi).  
\label{vcoup}
\end{equation}
Expanding the total wave function on the basis of intrinsic states
\begin{equation}
  \Psi = \sum_m \eta_m(X) \phi_m(\xi),
\label{eq:psi1}
\end{equation}
one obtains the following coupled equations to describe the 
macroscopic motion 
\begin{eqnarray}
  [- \frac{\hbar^2}{2\mu} \frac{d^2}{dX^2} +  U(X) -E]
  \left(
    \begin{array}{@{\,}c@{\,}}
      \eta_{1}(X) \\
      \eta_{2}(X)
    \end{array}
  \right) 
  +\left( 
    \begin{array}{@{\,}cc@{\,}}
      \epsilon_1 & 0 \\
      0 & \epsilon_2
    \end{array}
  \right) 
  \left( 
    \begin{array}{@{\,}c@{\,}}
      \eta_{1}(X) \\
      \eta_{2}(X)
    \end{array}
  \right)
  + f(X)
  \left( 
    \begin{array}{@{\,}cc@{\,}}
        M_{11} & M_{12} \\
        M_{21} & M_{22}
      \end{array}
  \right) 
  \left( 
    \begin{array}{@{\,}c@{\,}}
      \eta_{1}(X) \\
      \eta_{2}(X)
    \end{array}
  \right) =0. 
\label{cceq}
\end{eqnarray}
In the direct method, these coupled-equations are 
numerically solved with the following boundary conditions,
\begin{eqnarray}
\eta_n(X) &\rightarrow& t_n \frac{1}{\sqrt{k_n}} e^{-ik_nX} 
\hskip 3cm (X<<0)\\
&\rightarrow&
\frac{1}{\sqrt{k_n}}
[I_n e^{-ik_nX} + r_n e^{ ik_nX}], 
\hskip 1cm (X>>0)
\label{wfasym} 
\end{eqnarray}
where $k_n=\frac{\sqrt{2\mu(E-\epsilon_n)}}{\hbar}$. The amplitude of the 
incident wave $I_n$ is taken to be $\delta_{nn_0}$ if the intrinsic 
motion is initially in the state $\vert \phi_{n_0}>$. 
Once the transmission amplitudes 
$t_n$ are determined, the barrier transmission probability is given by,
\begin{eqnarray}
P=\sum_n \vert t_n \vert ^2 .
\label{pcc} 
\end{eqnarray}

In order to compare the results with those of the semi-classical 
mean field theory, the same problem can be cast into the form of the 
corresponding time dependent approach, where the wave function 
for the internal motion is expressed as 
\begin{equation}
  \label{eq:intwf}
  \phi(\xi,t)
  =a_1(t)\phi_1(\xi)+a_2(t)\phi_2(\xi)
  =a_1(t)
  \left(
    \begin{array}{@{\,}c@{\,}}
      1 \\
      0
    \end{array}
  \right) 
  + a_2(t)
  \left(
    \begin{array}{@{\,}c@{\,}}
      0 \\
      1
    \end{array}
  \right).  
\label{wfint}
\end{equation}
Following eqs.(\ref{envc}) and (\ref{envt}), 
the expansion coefficients $a_1$ and $a_2$ obey 
\begin{eqnarray}
  i\hbar \frac{d a_1(t)}{dt} &=& \epsilon_1 a_1(t) + f(X(t)) \{ M_{11} a_1(t) + M_{12} a_2(t) \}
\label{eq:a1}\\ 
  i\hbar \frac{d a_2(t)}{dt} &=& \epsilon_2 a_2(t) + f(X(t)) \{ M_{21} a_1(t) + M_{22} a_2(t) \},
\label{eq:a2}
\end{eqnarray}
in the classically allowed region, and 
\begin{eqnarray}
  -\hbar \frac{d a_1(\tau)}{d\tau} &=& \epsilon_1 a_1(\tau) + f(X(\tau)) \{ M_{11} a_1(\tau) + M_{12} a_2(\tau) \} 
\label{eq:a1t}\\
-\hbar \frac{d a_2(\tau)}{d\tau} &=& \epsilon_2 a_2(\tau) + f(X(\tau)) \{ M_{21} a_1(\tau) + M_{22} a_2(\tau) \},
\label{eq:a2t}
\end{eqnarray}
in the tunneling region. 
Using thus obtained $a_1$ and $a_2$ the renormalization of 
the potential barrier for the macroscopic motion by internal degrees 
of freedom is given by
\begin{equation}
  \Delta V(X(t))= \frac{\epsilon_1 |a_1(t)|^2 + \epsilon_2 |a_2(t)|^2
    + \sum^2_{m m'=1} M_{m m'} a_m^*(t)a_{m'}(t) \cdot f(X) }
  {|a_1(t)|^2 + |a_2(t)|^2} - \epsilon_0
\label{deltav}
\end{equation}
in the classically allowed region. 
The same expression holds for the tunneling region by 
replacing $a_i(t)$ with $a_i(\tau)$, i being 1 and 2.
Eqs.(\ref{eq:a1}) and (\ref{eq:a2}) or eqs.(\ref{eq:a1t}) and (\ref{eq:a2t}) 
should be solved together 
with eqs.(\ref{relc}) and (\ref{relt}), respectively,  
with the potential renormalization 
given by eq.(\ref{deltav}) in the classically allowed region 
and the corresponding one in the tunneling region 
to determine the classical trajectory $X(t)$ and $X(\tau)$.

One then calculates the barrier transmission probability 
for each initial kinetic energy $E_K$ by the WKB formula as
\begin{equation}
  \label{eq:penet}
  P(E)=exp(\frac{-4}{\hbar}\int_{\tau_b}^{\tau_a} d\tau 
  [U(X(\tau))+\Delta V(X(\tau))-E_K])
\label{psmft}
\end{equation}
where $\tau_a$ and $\tau_b$ correspond to the inner and outer classical 
turning points, respectively, and 
the time integral is performed over the tunneling region.

\section{Accuracy of the semiclassical mean field theory}

Before we discuss the characteristics of the intrinsic transition and the 
associated potential renormalization, we examine in this section 
the reliability of the semi-classical mean field theory for quantum tunneling.
We consider the case where 
the coupling matrix is given by
\begin{eqnarray}
  M=
  \left( 
    \begin{array}{@{\,}cc@{\,}}
      0 & 1 \\ 
      1 & 0
    \end{array}
  \right) 
\label{coupmat}
\end{eqnarray}
and assume that the bare potential and the coupling form factor have 
the Eckart form \cite{llqm}
\begin{equation}
  U(X)=\frac{U_0}{cosh^2(X/a)}, \ f(X)=\frac{f_0}{cosh((X-X_f)/a_f)},
\end{equation}
where the height of the potential barrier is fixed to be 
$U_0=10$ $MeV$, and 
the position of the coupling form factor
$X_f$ is taken to be either 50 $fm$ in front of the central region 
of the potential barrier or 0 $fm$. 
The large value of $X_f$ is taken, because we can then clearly 
separate the transition of the 
intrinsic state by the coupling form factor and by 
the tunneling effect. The latter transition 
and the associated potential renormalization,
i.e. the potential renormalization induced by 
the tunneling assisted intrinsic transition are the major issues 
which we wish to advocate in this paper. 
The two kinds of intrinsic transition mix up when $X_f=0$. 
The mass parameter is fixed to be $\mu c^2=2000 MeV$. 

The effects of intrinsic degrees of freedom are sensitive to the 
degree of adiabaticity measured by the 
adiabaticity parameter $\lambda=\frac{\omega}{\Omega}$, $\omega=\epsilon_2-
\epsilon_1$ and $\Omega$ being the excitation energy of the intrinsic motion 
and the curvature of the bare potential barrier $U(X)$
\cite{tak95,bt85,tak94,hag95}. 
We study a fast and slow tunneling represented by $\lambda=0.5$ and $2.0$, 
respectively. Since the whole tunneling process crucially depends on 
the initial condition of the intrinsic motion \cite{kt01}, 
we discuss both cases 
where the intrinsic motion is initially in the ground and in the excited 
states.

Fig.1 compares the barrier transmission probability as a 
function of the incident kinetic energy calculated in various ways 
for the case of $X_f=50 fm$. 
The upper and the lower panels correspond to 
$\lambda=0.5$ and $2.0$, respectively. They have been calculated by assuming 
$(a,a_f)=(5fm,5fm)$ and $(15fm,5fm)$, respectively. 
In both of them, the thick dashed line is the barrier transmission 
probability in the absence of the intrinsic motion. 
As it should, it becomes about 0.5 when the incident energy 
coincides with the barrier height 10 MeV.
The thick solid and the dot-dashed lines have been calculated 
by direct numerical integration of the coupled-channels equations,
i.e. by eq.(\ref{pcc}), 
for the cases where the intrinsic motion starts from the ground and 
the excited states, respectively. 
As expected, the barrier transmission probability is hindered 
over whole energy region 
by the intrinsic motion if it is initially in the ground state, 
while it is enhanced if the system starts from the excited state 
of the intrinsic motion. An interesting observation is that the 
fast intrinsic motion yields a step-function like 
structure in the excitation function of the barrier 
transmission probability,
while the barrier transmission probability shows no structure when 
the macroscopic variable couples with a slow intrinsic degree of freedom. 

In order to test the accuracy of the semiclassical mean 
field theory, the thin solid and the thick dotted lines 
have been calculated 
based on eq.(\ref{psmft}) 
for the cases where the intrinsic motion starts from the ground and 
the excited states, respectively. 
We observe that the semi-classical mean field theory well reproduces the 
results of the direct numerical integration of the coupled-channels 
equations at low energies, though it fails at high energies as 
a natural consequence of the well known fact that the simple 
WKB approximation should be replaced by the uniform approximation. 

We now consider the case when the coupling form factor is located 
in the same region as the bare potential barrier by setting 
$X_f=0 fm$.  The resultant barrier transmission probability is shown 
in Fig.2. Similarly to the case of Fig.1, the extension 
parameters have been 
chosen to be $(a,a_f)=(5fm,5fm)$ and $(15fm,5fm)$ for the cases of 
$\lambda=0.5$ and $2.0$, respectively. 
The notations are the same as those in Fig.1. 
As we discuss in \cite{kt01}, slow intrinsic motion corresponding to 
$\lambda=0.5$ enhances the 
barrier transmission probability at low energies and hinders it at 
high energies irrespective of whether the intrinsic motion 
is initially in the ground or in the excited states, though 
quantitative details are different. 
The fast intrinsic motion, on the other hand, enhances the barrier 
transmission probability when it starts from the ground state, and 
hinders when it starts from the excited state for a wide range of the 
incident energy. 
Similarly to the previous case when $X_f=50.0 fm$, the 
semiclassical mean filed theory well reproduces the results of the 
direct numerical integration of the coupled-channels equations 
at low energies.   
The accuracy of the semiclassical mean field theory can be more 
clearly seen in Fig.3, where the barrier transmission probability 
in low energy region is shown in an expanded scale. 
This figure is useful also to see 
that the hindrance effect of a fast intrinsic motion 
to the barrier transmission probability in the case 
when it starts from the excited state 
turns into an enhancement effect at low energies (see the lower panel).

\section{Properties of the intrinsic transition and the potential 
renormalization}

We now discuss the characteristics of the intrinsic transition and 
the associated potential renormalization. We first consider the case 
corresponding to Fig. 1, i.e. where the coupling form factor 
is located at far in front of the barrier region. 
Fig.4 shows the change of the occupation  
probability of the ground and the excited states, $\vert a(1) \vert^2$ 
(the thick solid line) and 
$\vert a(2) \vert^2$ (the thick dotted line), with the position X.  
The incident direction is toward the negative values of X. 
Similarly to Fig.1, 
the upper and lower panels correspond to the cases of coupling to 
slow or fast intrinsic motion. Each of them is divided into the left and 
right panels, where the results when the intrinsic state is initially 
the ground and the excited states, respectively, are shown. 
The bare potential $U(X)$ and the coupling form factor $f(X)$ are also 
shown in the figure in arbitrary scale. 
In the tunneling region, 
$\vert a(1) \vert^2$ and $\vert a(2) \vert^2$ 
have been normalized such that 
$\vert a(1) \vert^2$ + $\vert a(2) \vert^2$ = 1.0. 
The behaviour of $\vert a(1) \vert^2$ and $\vert a(2) \vert^2$ 
as functions of $X$ depends on the incident energy. By considering the 
accuracy of the semiclassical mean field theory discussed in the 
previous section, 
we chose $E_K$ to be 
8.74, 7.64, 9.4 and 6.87 $MeV$ 
to obtain the upper left, upper right, the lower left and lower right 
panels, respectively. 
The semiclassical mean field theory well reproduces the barrier 
transmission probability obtained by the direct numerical solution 
of the coupled-channels equations at these energies. 
In the figure, the short horizontal line in the barrier region 
indicates the tunneling region. 

The behaviour of the occupation probability can be easily 
understood in the present case. 
The intrinsic degree of freedom makes transition to the other state 
from the initial state 
when the macroscopic motion traverses the coupling region. 
The interesting 
result is that it then makes a strong transition to the ground state 
in the tunneling region even though the coupling form factor is already 
almost zero in that region. 
We name this the tunneling induced quantum transition of the 
intrinsic motion, 
because these transitions take place to increase the tunneling probability. 

We now calculate the potential renormalization and the 
effective potential barrier based on eq.(\ref{deltav}) for the classically 
allowed region and the corresponding expression for the tunneling region. 
The results are shown in Fig.5. The input parameters are 
the same as those for Fig.4. The figure contains the 
potential renormalization $\Delta V(X)$ in the case 
when the intrinsic motion starts from the 
ground (the solid line) and the excited states (the thick dotts), 
the effective potential $U_{eff}(X)=U(X)+\Delta V(X)$ 
for these cases (the dotts and the dott-dashed 
lines). The coupling form factor is also shown (the thin dashed line). 
We see two different kinds of potential renormalization in 
Fig.5. The first is the potential renormalization in the region of the 
coupling form factor, which one naturally expects to exist. 
The other is the novel type of potential renormalization that is caused 
by the tunneling induced intrinsic transition. The important 
consequence is that the effects of intrinsic motion 
cannot be represented by a constant potential shift 
in the tunneling region even though the 
coupling form factor is outside the tunneling region. 

Let us now consider the case when $X_f=0.0 fm$. 
The resultant 
position dependence of the occupation probabilities 
$\vert a(1) \vert^2$ and $\vert a(2) \vert^2$  and the associated 
potential renormalization and the net effective potentials are 
shown in Figs. 6 and 7, respectively. The meaning of each line is the same 
as that in Figs. 4 and 5.
The incident kinetic energy has been chosen to be 
6.93, 5.0, 8.96 and 6.43 $MeV$ 
to obtain the upper left, upper right, the lower left and lower right 
panels, respectively, in Fig.6. 

Since the two effects, i.e. the transition by the coupling form factor and 
that induced by the tunneling effect, overlap, 
the behaviour of $\vert a(1) \vert^2$ and $\vert a(2) \vert^2$ 
in the present case gets more complex than the previous case, 
where the coupling form factor is far outside the tunneling region. 
The important thing is that 
the potential renormalization is strongly spatially dependent 
also in this case. 
Another interesting observation is that the potential renormalization 
is larger in the case where the intrinsic motion starts from the 
excited state than the case when it starts from the ground state. 
This is related to the choice of the incident energy to draw 
Fig. 6 and is reflected in the larger enhancement of the barrier transmission 
probability in the former case as is shown in Fig.3.

\section{Summary}

The effect of environmental or intrinsic degrees of freedom 
on the tunneling probability of a macroscopic variable has been 
one of the very active subjects in many fields of physics and chemistry 
in the past decades \cite{cl,ha95,weiss}. 
One interesting question in this respect is to understand 
the way the intrinsic degrees of freedom modify the tunneling potential 
barrier. This question is especially important in connection with 
low energy nuclear reactions in laboratories, because the large 
deviation of the experimental reaction rate from the extrapolation 
from high energy data has been analyzed by assuming a constant potential 
shift in the tunneling region, which is supposed to well 
represent the screening effects by bound target electrons, 
with a puzzling consequence that the obtained 
screening energy exceeds the theoretically expected upper limit, i.e. 
the adiabatic limit, for all systems so far studied. 

In this paper, we have set a simple two-level coupled-channels problem, and 
analyzed the properties of the intrinsic transition and the 
associated potential renormalization by using a semi-classical 
mean field theory for quantum tunneling, which naturally 
introduces the concept of the potential renormalization. 

One of the main conclusions of our study is that there can occur 
strong intrinsic transitions in the tunneling region in order 
to facilitate the quantum tunneling even if there is 
no coupling any more in that region. 
We called this phenomenon the tunneling induced or assisted 
intrinsic transition. We have shown that it leads to a strongly 
spatially dependent potential renormalization in the tunneling region.   
This suggests that 
the screening effects in realistic low energy nuclear reactions should be 
handled not by simply assuming a constant potential shift as has been done 
in all previous studies, but by a proper dynamical treatment of the 
tunneling region. We publish our study in this direction in  
separate papers \cite{ktab01,kbt01}.

\bigskip

\noindent
Acknowledgment The authors wish to thank 
Bertrand Giraud for useful discussions.
This work has been partly accomplished while 
the authors stayed at the INFN at Catania. 
They thank M. Di Toro, A. Bonasera, C. Spitaleri and S. Terranova 
for useful discussions and warm hospitality. 
This work is supported by
the Grand-in-Aid for Scientific Research from 
Ministry of Education, Culture, Sports, Science and Technology 
under Grant No. 12047203 and No. 13640253
and by the Japan Society for the Promotion 
of Science for Young Scientists under the contract No. 12006231.


\newpage

\begin{figure}[l]
\includegraphics{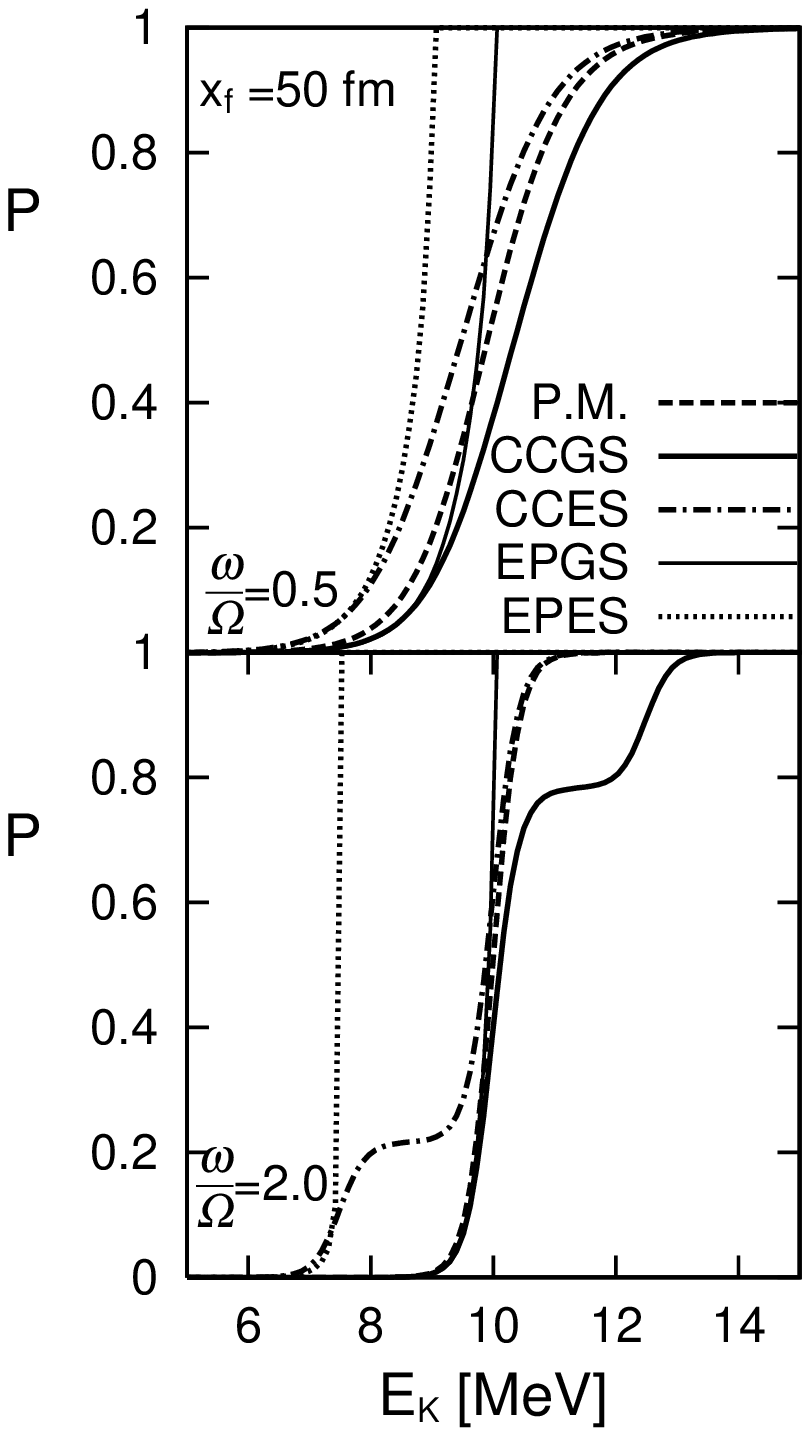}
\caption{\label{fig:fig01} The barrier transmission probability as a function of the incident kinetic energy for $X_f=50 fm $ calculated in various ways.} 
\end{figure}

\begin{figure}[l]
\includegraphics{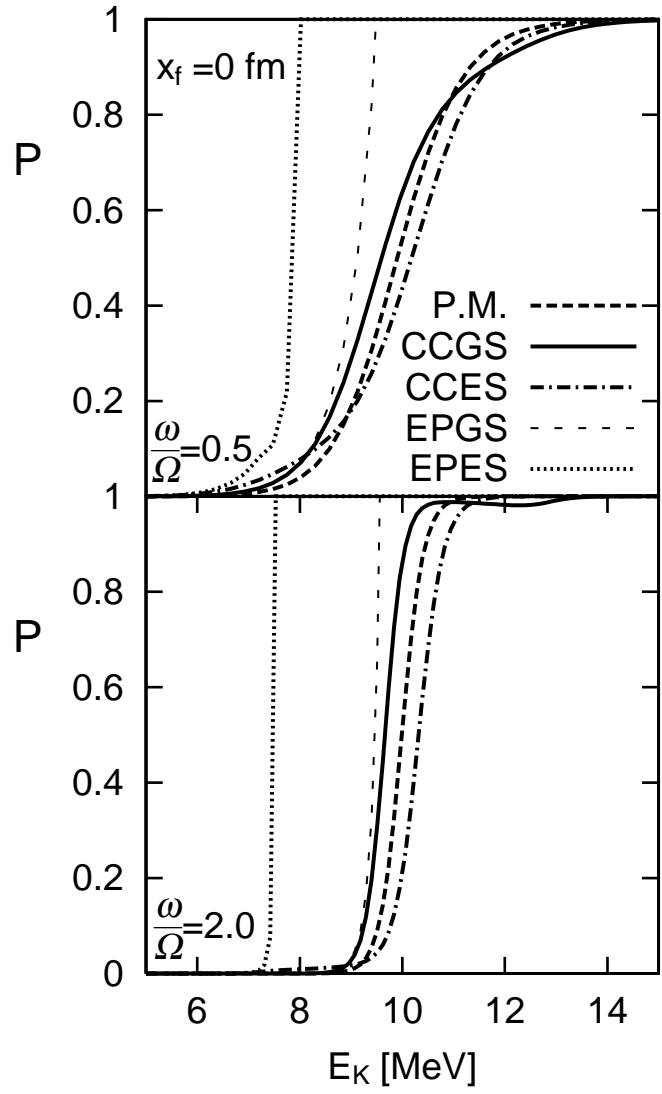}
\caption{\label{fig:fig02} The same as Fig.1, but for $X_f=0 fm $.}
\end{figure}

\begin{figure}[l]
\includegraphics{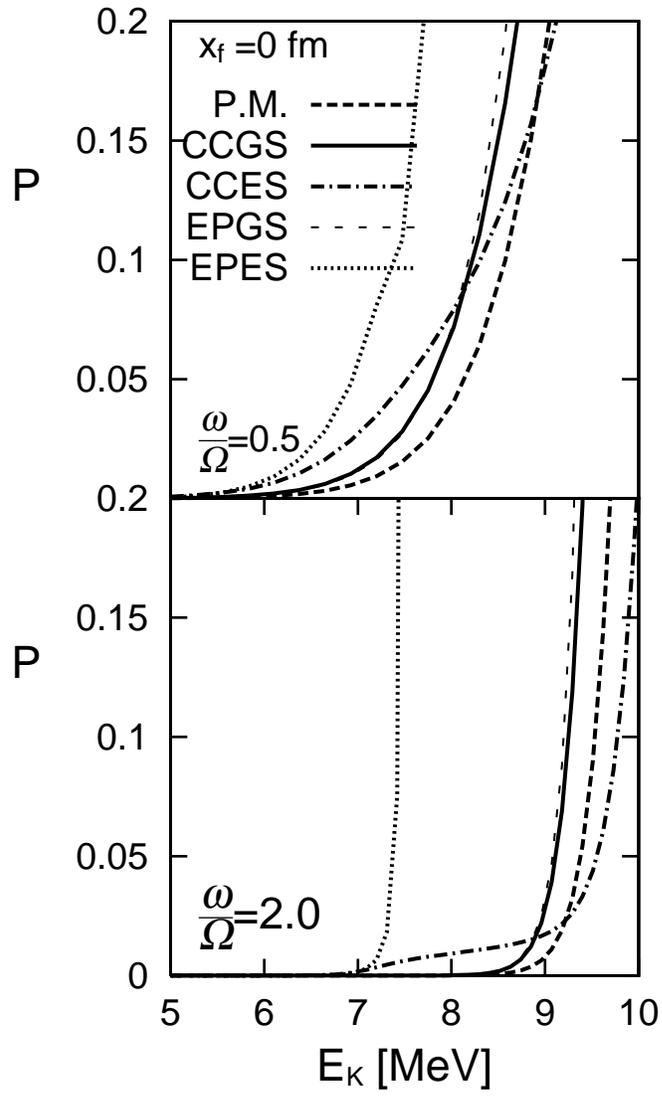}
\caption{\label{fig:fig03} The same as Fig.2, but in an expanded scale.}
\end{figure}

\begin{figure}[l]
\includegraphics{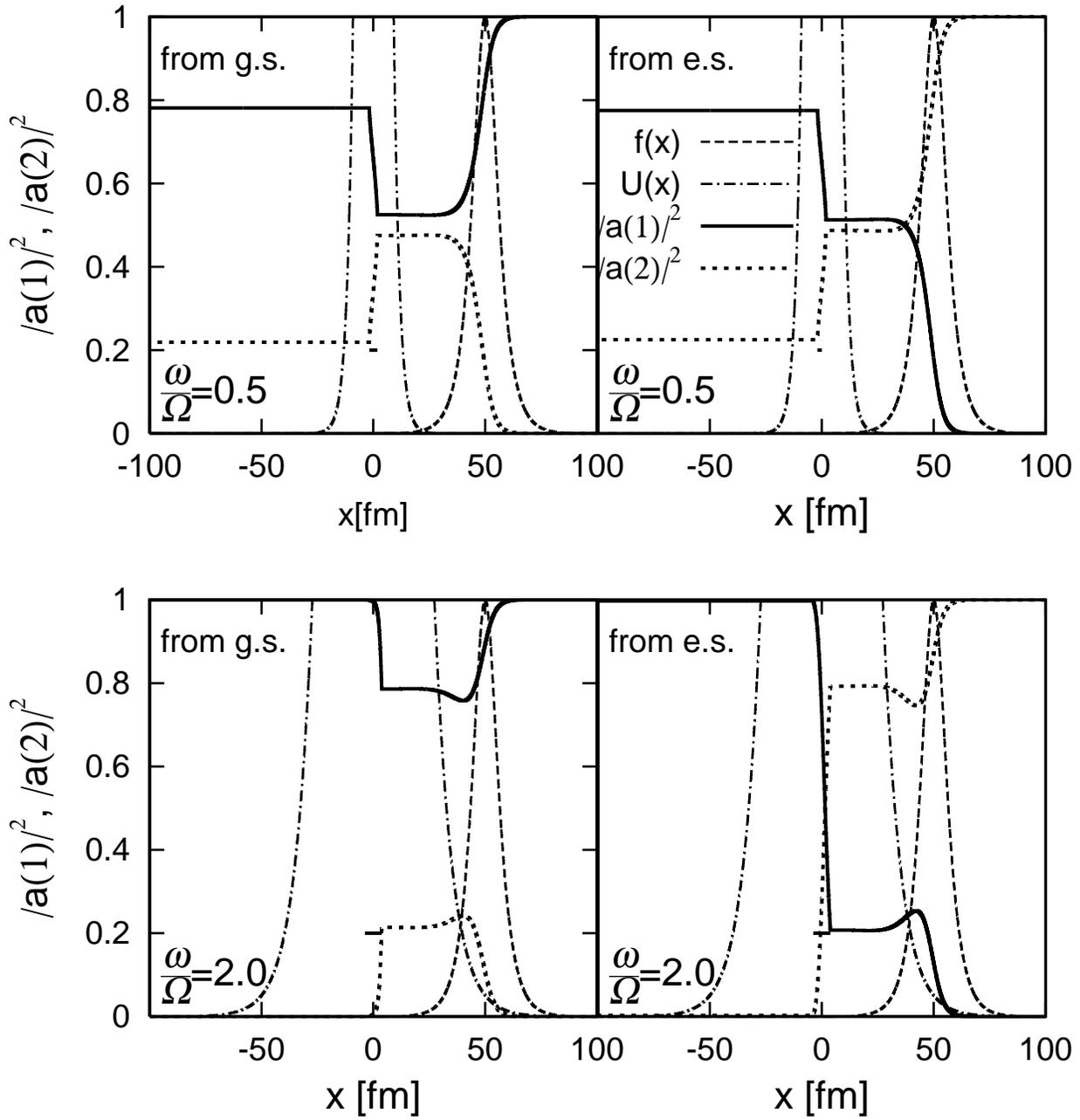}
\caption{\label{fig:fig04} The change of the occupation probabilities of the ground and the excited states of the intrinsic motion as functions of the position $X$ for $X_f=50 fm$ corresponding to the case shown in Fig.1. The incident energy for each of the four panels is given in the text.}
\end{figure}

\begin{figure}[l]
\includegraphics{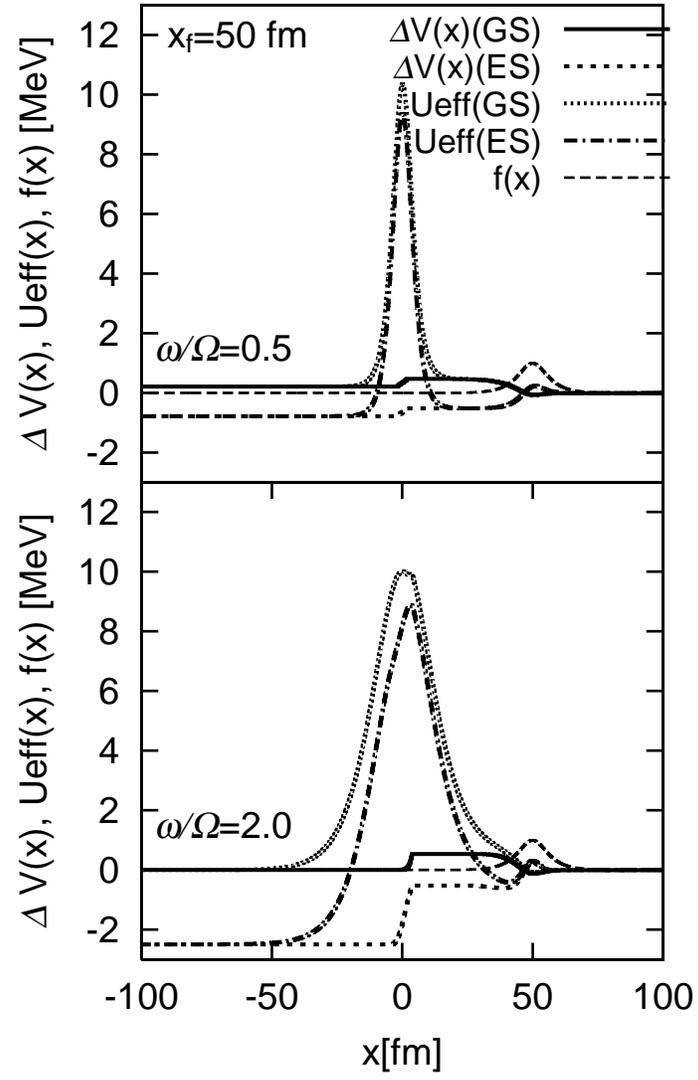}
\caption{\label{fig:fig05} The potential renormalization and the effective potential corresponding to each of the four cases shown in Fig.4.}
\end{figure}

\begin{figure}[l]
\includegraphics{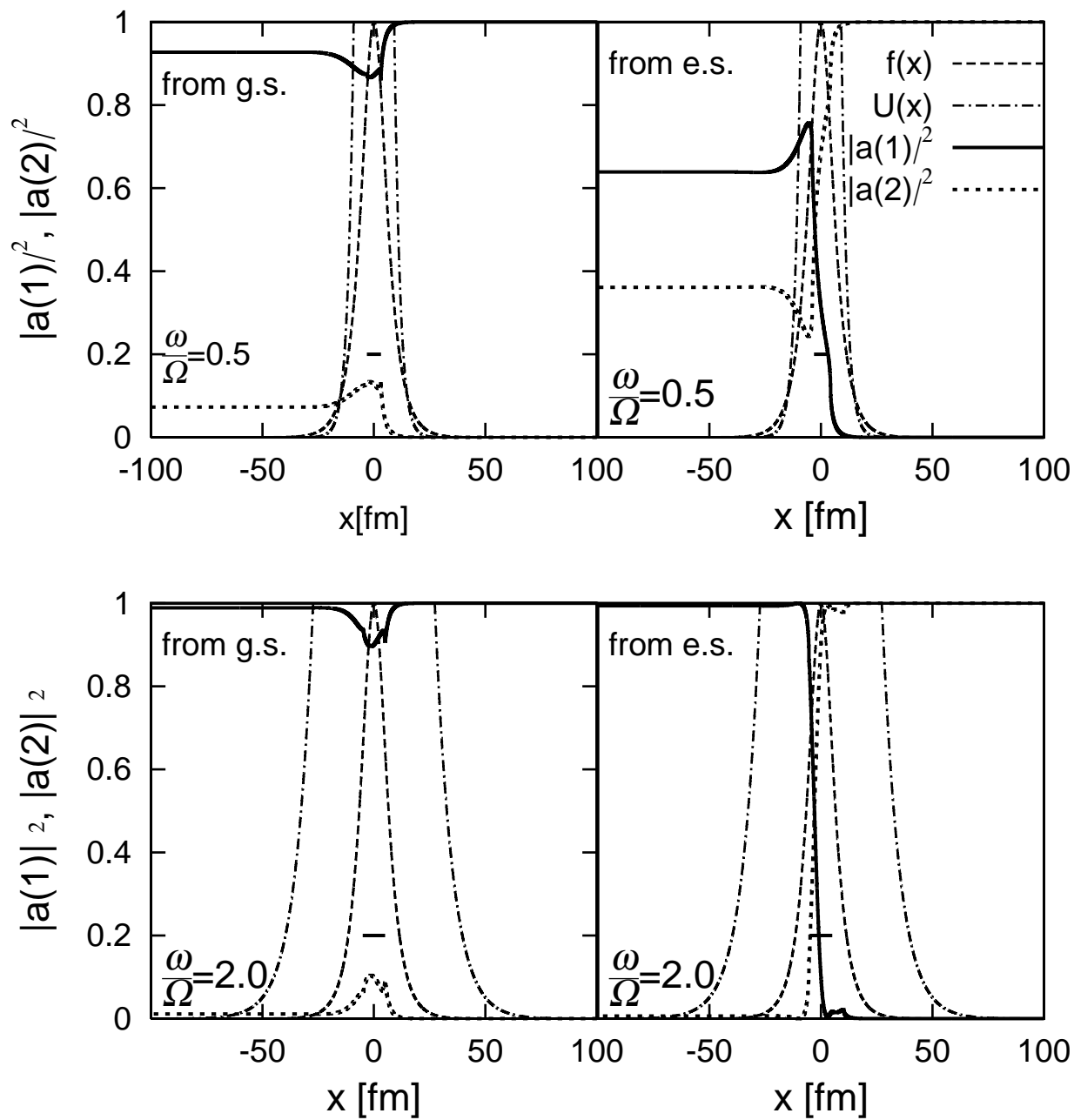}
\caption{\label{fig:fig06} The same as Fig.4, but for $X_f=0 fm$ corresponding to 
Figs.2 and 3.}
\end{figure}

\begin{figure}[l]
\includegraphics{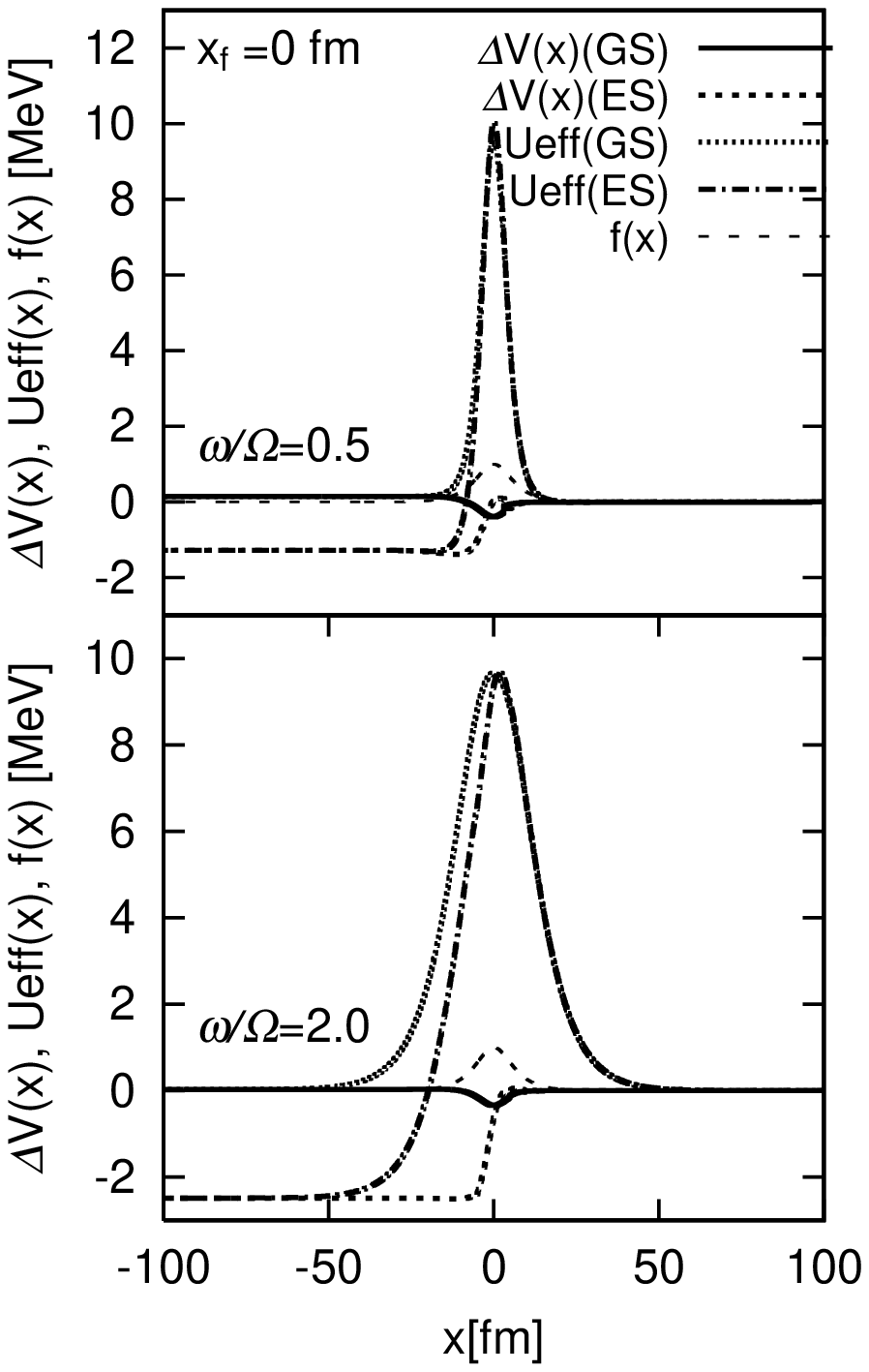}
\caption{\label{fig:fig07} The same as Fig.5, but for $X_f=0 fm$ corresponding to 
Fig.6.}
\end{figure}

\end{document}